\documentclass[12pt]{article}
\usepackage{amsmath,float,amssymb,amsthm,amsxtra,overpic,bbm,bm,epsfig,color}

\def\thefootnote{\fnsymbol{footnote}}

\def\tabnotefont{\fontsize{9}{10}\selectfont}%
\newenvironment{tabnote}{\par\tabnotefont}{\par}

\usepackage{url}
\usepackage{slashed,stmaryrd}
\usepackage{color}

\usepackage[numbers]{natbib}
\usepackage[colorlinks=true,citecolor=blue,linkcolor=blue]{hyperref}
\usepackage{amsmath,amssymb}
\usepackage{graphicx,color}
\usepackage{sidecap}
\usepackage{footnote}
\makesavenoteenv{tabular}
\usepackage{xspace}
\usepackage{makecell}

\newcommand{\Thorium}     {$^{232}$Th\xspace}
\newcommand{\Uranium}     {$^{238}$U\xspace}
\newcommand{\tracerU}     {$^{233}$U\xspace}
\newcommand{\tracerTh}    {$^{229}$Th\xspace}
\newcommand{\nitric}      {HNO$_3$\xspace}
\newcommand{\ammoniaW}    {$\rm NH_3\cdot H_2O$\xspace}
\newcommand{\nitricZr}    {$\rm Zr(NO_3)_4$\xspace}
\newcommand{\nitricCu}    {$\rm Cu(NO_3)_2$\xspace}

\newcommand{\molperl}     {mol$\cdot$L$^{-1}$\xspace}
\newcommand{\FeCl}        {FeCl$_3$\xspace}

\newcommand{\ZrCl}        {ZrCl$_4$\xspace}

\begin{document}
\vspace{0.2cm}

\begin{center}
{\Large\bf Co-precipitation approach to measure amount of \Uranium in copper to sub-ppt level using ICP-MS}
\end{center}

\vspace{0.2cm}

\begin{center}
{\bf Ya-Yun Ding $^{a}$}, \footnote{E-mail: dingyy@ihep.ac.cn}
\quad
{\bf Meng-Chao Liu $^{a}$},
\quad
{\bf Jie Zhao $^{a,b}$},
\quad
{\bf Wen-Qi Yan $^{a}$},
\quad
{\bf Liang-Hong Wei $^{a}$},
\quad
{\bf Zhi-Yong Zhang $^{a}$},
\quad
{\bf Liang-Jian Wen $^{a,b}$}, \footnote{E-mail: wenlj@ihep.ac.cn}
\\
{\small $^a$Institute of High Energy Physics, Chinese Academy of Sciences, Beijing 100049, China}\\
{\small $^b$State Key Laboratory of Particle Detection and Electronics, Institute of High Energy Physics, CAS, Beijing 100049, China}
\end{center}

\vspace{1.5cm}

\begin{abstract}
Inductively coupled plasma mass (ICP-MS) spectroscopy is widely
used for screening materials of low background detectors in dark
matter and double beta decay searches due to its high sensitivity to
trace \Uranium and \Thorium.
This work describes a novel co-precipitation approach to
measure the amount of \Uranium in high-purity copper to sub-ppt level.
Such an approach allows the pre-concentration of U and removal of
the matrix, by selecting a proper precipitator to co-precipitate with
\Uranium and using excess ammonia water to separate the uranium hydroxide
from copper by forming water-soluble tetra-amminecopper (II).
The isotope dilution method and standard addition method were both
used to mitigate the matrix effect and cross-check each other.
The latter was also used to measure the recovery efficiency of \Uranium
by using \tracerU as the tracer.
The method detection limit (MDL) reached $\sim$0.1 pg \Uranium /g Cu
for both methods while the recovery efficiency of uranium robustly
remains 65\%--85\%.
Various sources of interference in the ICP-MS analysis were thoroughly evaluated,
and the contamination from reagents were found to be the dominant factor
that affected the MDL. Further purification will allow
significant improvements in the MDL.
This co-precipitate approach can be easily extended to measure \Thorium
by using \tracerTh as the tracer.
\end{abstract}

\begin{flushleft}
\hspace{0.9cm} Keywords: ultra-low radioactivity, ICP-MS, co-precipitation, uranium
\end{flushleft}

\def\thefootnote{\arabic{footnote}}
\setcounter{footnote}{0}

\newpage

\section{Introduction}
\label{sec:introduction}

Experiments searching for neutrino-less double beta decay (NDBD)
have a stringent requirement on the natural radioactivity in the
detector materials, particularly \Uranium and \Thorium.
The capability to detect ultra-trace amounts of \Uranium and
\Thorium allows careful assessment of the backgrounds,
and thus is important to the planned NDBD
experiments~\cite{Albert:2017hjq,Kharusi:2018eqi,Abgrall:2017syy,Chen:2016qcd}.
In addition to NDBD experiments, many neutrino experiments
or dark matter experiments also require ultra-low-radioactivity
materials for the detector construction.
The highly sensitive methodologies known to date include
neutron activation analysis and inductively coupled plasma
mass spectroscopy (ICP-MS)~\cite{Leonard:2007uv}.
ICP-MS analysis is relatively quick and typically has an intrinsic
detection limit better than parts-per-trillion (ppt or pg/g) for
a large number of elements, particularly uranium and thorium.

ICP-MS analysis is optimized for aqueous samples, e.g., it is
commonly used to measure uranium and thorium in water samples
~\cite{Rozmaric2009, Takata2011}.
However, the drawback of using this technique to detect ultra-trace
elements is the need for arduous pre-treatment of the samples.
Acid digestion is typically used to deal with the
solid materials, and the tolerable total
dissolved salts (TDSs) in the matrix for ICP-MS analysis is $<0.1$\%.
The pre-treatment inevitably needs additional reagents
and chemical separation processing to reduce TDSs or concentrated
U and Th. It may consequently introduce contamination and
result in worse detection
limits~\cite{Unsworth2001,Pin2001,Pollington2001}.
Thus the pre-treatment approach must be scrupulously
designed and carried out.

Electrochemical techniques, precipitation, and ion exchange are
widely used methods used to separate the matrix and analyte~\cite{RAO2006}.
Recently, an anion exchange method was reported
in~\cite{LaFerriere:2014rva} that achieved detection limits of
10$^{-2}$ pg/g level for both \Uranium and \Thorium in copper samples.
With anion exchange, this method effectively
concentrated analytes and simultaneously removed an unwanted sample matrix.
The radio-assay to ultra-trace U and Th is extremely challenging.
Development of other convenient methods with different pre-treatments
is highly desired and would benefit future experiments.
It is well known that Cu has the complex ability to form water-soluble
compounds in excess ammonia water, whereas U will precipitate.
However, ultra-trace U or Th in solutions are too diluted to be
separated by conventional methods. The co-precipitation approach is
feasible in this case. Furthermore, to achieve a sub-ppt-level
detection limit, it is critical to scrupulously perform sample
preparation, and qualify the cleanness of labware and the purity of reagents.
Calibration of the recovery efficiencies of U or Th during the pre-treatment
is also essential, and the typical way is to use \tracerU or \tracerTh
as a tracer. We have obtained \tracerU and \tracerTh
standard solutions, but only the former is qualified to have sufficiently
low contamination with respect to the target element.
Hence, in this work, we present the \Uranium measurement to demonstrate
this approach, but it should be pointed out that the method is also applicable
to the \Thorium measurement.

The rest of this paper is organized as follows:
In Sec.~\ref{sec:exp}, the experimental details to detect the
ultra-trace amount of \Uranium in copper is introduced, highlighting the
pre-treatment process with a co-precipitator and tracer. The various
sources of interference in the ICP-MS analysis are discussed in Sec.~\ref{sec:opt}.
Two separate analyses with the standard addition method and the isotope
dilution method are presented in Sec.~\ref{sec:result},
and they give a consistent detection limit.
Finally, we summarize this work and discuss its prospects
in Sec.~\ref{sec:summary}.

\section{Experimental Section}
\label{sec:exp}

\subsection{Instruments, reagents, and labware}

All chemical operations and measurements were done in a class 10,000
clean room to suppress environmental contamination, since the
concentration of \Thorium and \Uranium in dust is typically
around the ppm level. The analyses were performed using a ThermoFisher
iCAP-Qc Quadrupole ICP-MS instrument with a PFA concentric nebulizer.
A collision cell was equipped but not used in this work.
The spectrometer was tuned every few months.
For 1 pg \Thorium or \Uranium per gram solution, the typical
counts per second with the ICP-MS device is approximately 700--900.
Before measuring the samples, a standard \Uranium solution with
a ppt-level concentration should be tested for at least 30
min to ensure that the instrument is stable. Between every two samples,
ultra-pure \nitric solution was measured to ensure that the entire
system was clean.

The reagents used in this work are listed in Table~\ref{tab:reagents}.
The water for cleaning and dilution was unexceptionally ultra-pure water.
The ammonia water was further purified before use.
The \ZrCl and \FeCl, both of 99.99\% purity,
were dissolved in ultra-pure water.
The \tracerU standard solution contains
2.3539(42)$\times$10$^{-6}$ g (\tracerU)/g (solution).
The \Uranium standard solution was diluted with 5\% \nitric.

\begin{table}[!htb]
\centering
\caption{Reagents used in this work. \label{tab:reagents}}
\begin{tabular}{cccc}
\hline
Reagents or Labware & Description \\
\hline
ultra-pure water &  Milli-Q$^\circledR$ Reference   \\
nitric acid   &  OPTIMA (Fisher Scientific)     \\
ammonia water  &  BV-III (BICR $^a$)   \\
\ZrCl, \FeCl  &  Sigma-Aldrich       \\
\tracerU standard  &  IRMM $^b$      \\
\Uranium standard &   100 $\mu$g$\cdot$mL$^{-1}$  $^c$   \\
\hline
\end{tabular}
\begin{tabnote}
  $^{\rm a}$ BICR refers to Beijing Institute of Chemical Reagents.\\
  $^{\rm b}$ IRMM refers to Institute for Reference Materials and Measurements.\\
  $^{\rm c}$ From national standard reference material of P.R. China.
\end{tabnote}
\end{table}

The filter units were Millex-LG$^\circledR$ sterilizing filter
units (Millipore Ireland BV, Carrigtwohill, Co. Cork),
which were tested to be compatible with 6 \molperl \nitric.
The 20-mL polypropylene syringes were produced in JiangSu, China.
All vials, containers, pipette tips, stirring bars, filter units,
and syringes were cleaned with Alconox$^\circledR$ detergent and
rinsed with ultra-pure water at least three times, followed by at
least two overnight leaches in 6 \molperl electronic grade \nitric,
followed by rinsing with water. All labware was filled with or immersed
in hot 6 \molperl OPTIMA \nitric for 20 min prior to use,
followed by at least three rinses with water.

\subsection{Pre-treatment process}
\label{sec:samplePrep}

\begin{figure}[!htb]
\centering
\includegraphics[width=0.9\columnwidth]{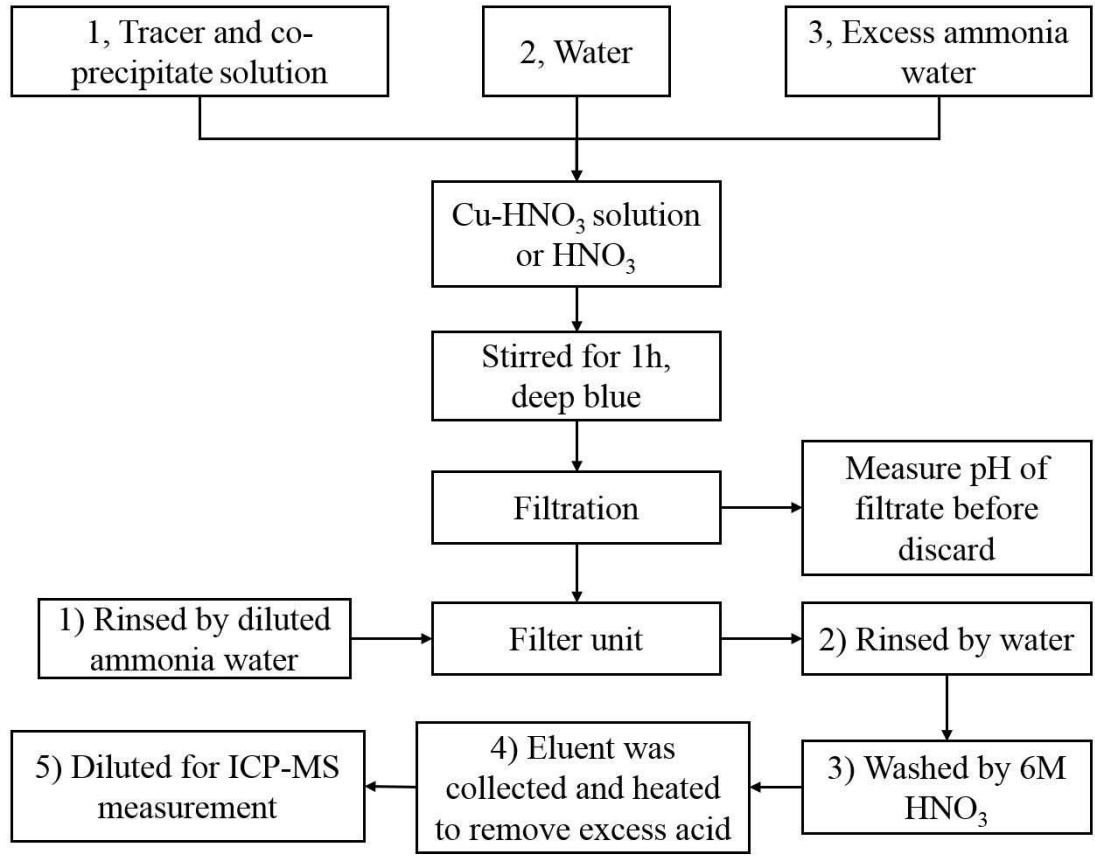}
\caption{Flowchart of pre-treatment process. } \label{fig:flow}
\end{figure}

The pre-treatment flow is shown in Fig.~\ref{fig:flow}.
The key idea is to form water-soluble Cu ammonia complex with
excess \ammoniaW, which allows the separation between the bulky Cu
and the \Uranium and \Thorium elements that precipitate in such
circumstances. The major steps include dissolution, co-precipitation,
and filtration. During pre-treatment, the calibration of recovery
efficiency is critical and also implemented.

\subsubsection{Dissolution}

The copper samples were first soaked in the detergent
(Alconox$^\circledR$) aqueous solution for 15 min under
ultrasonic conditions, then rinsed and soaked in fresh water for
5 min three times under ultrasonic conditions. Finally
they were dried with high-purity nitrogen.
To remove the surface contaminants, the cleaned copper was etched in
fresh 6 \molperl \nitric twice until its total weight lost was
$\sim$3\% by mass. The remaining copper was rinsed
with water, dried with nitrogen gas,
and completely dissolved by using 8 \molperl \nitric.
A reddish-brown smoke was visible in the beginning of dissolution,
which indicates the formation of NO$_2$:
\begin{equation*}
\rm Cu + 4 HNO_3 = Cu(NO_3)_2 + 2 NO_2\uparrow + 2 H_2O.
\end{equation*}
As the acid is consumed over time, the reaction equation becomes:
\begin{equation*}
\rm 3 Cu + 8 HNO_3 = 3 Cu(NO_3)_2 + 2 NO\uparrow + 4 H_2O.
\end{equation*}
The molar ration of Cu to \nitric was chosen to be 1:4 for complete
dissolution, and the excess \nitric helped to maintain the stability
of the copper solution.

\subsubsection{Recovery efficiency calibration}
\label{sec:recovery}

Recovery efficiency is an important indicator used to evaluate the
effectiveness of pre-treatment and must be obtained.
To evaluate the \Uranium recovery efficiency during
the pre-treatment process, \tracerU can be used as a tracer.
Because \tracerU does not naturally exist, its recovery efficiency
can be determined easily. As an isotope of uranium, \tracerU has
the most similar chemical properties as \Uranium.
If the \tracerU content is at the same level as \Uranium
in the solution, their recovery efficiencies are expected to be
the same. In addition, it is possible to use the isotope dilution
method with \tracerU.

Our tests showed that when \Uranium in aqueous solution is at
10$^{-9}$ g/g (ppb) level, the \Uranium recovery efficiency can be
$>$90\% by using the normal precipitation method with ammonia water.
However, if the \Uranium content is at
10$^{-12}$ g/g (ppt) level or lower, the \Uranium recovery
efficiency decreases to almost zero, and thus the co-precipitator
must be introduced.

\subsubsection{Co-precipitation}

The co-precipitation approach is widely used in radiochemistry
to separate elements that are too diluted to be
separated by conventional methods.
All metal elements that do not precipitate in ammonia water,
such as Ag, Cu, Zn and Mg, cannot be chosen as co-precipitator.
Fe and Zr ions are found to be more suitable candidates.
The \tracerU recovery efficiency for blanks (see Sec.~\ref{sec:blank})
was measured to be 71\%$\pm$4\% with \ZrCl as co-precipitator
and 48\%$\pm$2\% with \FeCl as co-precipitator, respectively;
thus, \ZrCl was chosen for this assay.

As shown in Fig.~\ref{fig:flow}, the quantitative \tracerU
standard solution and \ZrCl solution are first added into the
\nitricCu solution. Then, water is added in advance to dissolve
the copper ammonia complex formed later. Third, the ammonia water
is added slowly under stirring and the Cu$^{2+}$ ions start to precipitate:
\begin{equation*}
\rm [Cu(H_2O)_6]^{2+} + 2 OH^- \rightarrow Cu(OH)_2\downarrow+ 6 H_2O.
\end{equation*}
The ions of elemental Th, U, and Zr precipitate as well.
When adding excess ammonia water to the \nitricCu solution,
the $\rm Cu(OH)_2$ precipitates start to disappear gradually:
\begin{equation*}
\rm Cu(OH)_2 + 4 NH_3 + 2 H_2O \rightarrow [Cu(NH_3)_4 (H_2O)_2]^{2+} + 2 OH^{-}.
\end{equation*}
The solution turns dark blue due to the formation of
water-soluble tetra-amminecopper (II) in excess ammonia water,
whereas the precipitates due to Th, U, and Zr ions still exist.
In this process, one molar Cu needs four molar \nitric and six
molar \ammoniaW. More ammonia water and \ZrCl may improve the
recovery of metal ions, but the impurities in these two reagents
will degrade the detection limits of ICP-MS analysis.
Thus, less ammonia water or \ZrCl is preferred.

\subsubsection{Filtration}

The mixture after co-precipitation can be separated by filtration
with syringe-operated filter units.
The trace amount of the precipitates from U ions will be adsorbed
on the zirconium hydroxide and intercepted by the filter.
Then, the precipitates inside the filter unit are washed by warm
6 \molperl \nitric after rinsing with a small amount of diluted ammonia
water and pure water.
The eluent is collected and heated to remove excess acid until
the residual liquid is less than 0.4 g. The last step is to dilute
the residual liquid with 5\% \nitric to a certain amount.
The final solution is ready for ICP-MS analysis, and called
{\it Cu-sample} in the following context.

\subsubsection{Blanks}
\label{sec:blank}

Instrumental analysis is essential to determine the detection
limit when any pre-treatment is involved. The typical way is to
prepare the blank solution (called {\it blank} in the following context),
which should be prepared by following exactly the same procedure as
that for the sample, but without the sample.
For instance, the starter solution for the Cu-sample
is a certain amount of \nitricCu solution; thus, the starter
solution for the blank is the equivalent amount of nitric
acid to make that \nitricCu solution.
The contamination introduced from air, vessels,
water, and reagents can be estimated by blank measurements.

\subsubsection{Pre-treatment optimization}

For blanks, an equivalent amount of \nitric for dissolving Cu is
used to replace the Cu-solution.
The main reaction occurring in the blanks is:
\begin{equation*}
\rm HNO_3 + NH_3\cdot H_2O \rightarrow NH_4NO_3 + H_2O.
\end{equation*}
Different molar rations of nitric acid to ammonia water,
$n_{\rm HNO_3}:n_{\rm NH_3H_2O} = $4:5.8, 4:6, 4:6.4, 4:6.6, 4:11.5,
were tested while keeping the Zr concentration 3.5 ppm after
adding all chemicals. The Cu samples were tested under the same Zr
concentration, and the molar ratio of Cu to ammonia water was
chosen to be 1:5.8 and 1:6.4, respectively.
No obvious changes were observed in the \tracerU recovery efficiencies
with different pH values.
Thus, the molar ration $n_{\rm Cu}: n_{\rm HNO_3}: n_{\rm NH_3H_2O} =1:4:5.8$
was chosen to dissolve Cu and form Cu ammine complex.
The pH value of the mixture after adding ammonia water and stirring
was monitored, and it was 8.2--9.0 for blanks and 8.5--8.9 for Cu-samples.

The \ZrCl concentration was optimized accordingly.
For blanks, when the Zr concentration during pre-treatment increased
from 1.7 ppm to 3.5 ppm, the \tracerU recovery efficiencies were
measured to be 70\%$\pm$3\% and 65\%$\pm$2\%, respectively.
For Cu-samples, when Zr concentration was 1.7 ppm,
the \tracerU recovery efficiency was measured to be 64 $\pm$5\%.
Finally, the Zr concentration for the pre-treatment was chosen to
be 1.7 ppm, resulting in 0.026\% \nitricZr left in the blank or Cu-sample.

The reaction between metal ions and \ammoniaW was quick.
After adding \ammoniaW, the freshly formed Zr precipitate was left
in the solution under stirring to form larger particles,
which benefits the following filtration step.
Different stirring times from 1 to 3 h were tested, and
no difference was observed in the measured \tracerU recovery
efficiencies.

\section{Interference analysis}
\label{sec:opt}

After pre-treatment, the zirconium nitrate will remain in the solution
for both blanks and Cu-samples.
Part of the impurities from Cu or \ZrCl will also remain in the solution.
It is important to evaluate and eliminate the interference
to reduce the error. Typically the sources of interferences for
ICP-MS analysis include non-mass spectroscopic interferences,
spectral interferences, and contamination.

\subsection{Non-mass spectroscopic interference}
\label{sec:specInterference}

Non-mass spectral interference is also called physical
interference, and the most critical one in this work is the matrix
effect. For blanks, the main dissolved salts in the solution are
\nitricZr and other metallic nitrates formed by impurities in \ZrCl.
For Cu-samples, the main dissolved salts are almost the same as
the blanks, except for extra metallic nitrates originating from
impurities in the copper.
The \nitricZr content was estimated to be 0.026\% by mass.
The content of Cu$^{2+}$ ions in the Cu-sample was measured to be
approximately 5$\times$10$^{-5}$ g/g, or equivalently 0.015\%
\nitricCu by mass.
The maximum concentration of metal ions originating from Cu
impurities was estimated to be 3$\times10^{-5}$ by mass
when the purity of Cu is 99.995\%.
Other metal salts originating from \ZrCl in Cu-samples and blanks
were estimated to be 10$^{-6}--10^{-7}$ by mass.
Thus the TDSs in blanks or Cu-samples were
$<$0.1\% by mass, which is close to the upper limit of
TDSs for ICP-MS measurement~\cite{Beauchemin2010}.
This can be ignored for ordinary analysis, but must be taken
into account for the determination of ultra-trace
concentrations of analytes.

There are several ways to mitigate or eliminate the matrix effect,
such as the internal standard method, standard addition method,
argon gas dilution, and chemical separation method.
As one of the internal standard methods, the isotopic dilution
method allows a relative measurement by adding a known amount
of standard (isotopically enriched form of the analyte) to the
sample and measuring the ratio of analyte to standard.
This can significantly suppress the matrix effect since the
analyte and standard are in the same matrix.
However, the \tracerU recovery efficiency after pre-treatment
cannot be obtained by using the isotopic dilution method.
To obtain the recovery efficiency and cross-check it,
the standard addition method can be used to simultaneously
determine the efficiencies for both \tracerU and \Uranium,
because the matrix effect for both elements are perfectly matched.
The analysis details of these two methods are described
in Sec.~\ref{sec:result}.

\subsection{Spectral interference}

Sources of spectral interference include polyatomic interference and
isobaric interference. Evaluation of the spectral interference
is necessary in this work due to the ultra-low concentration
of \Uranium and the relatively high content of dissolved salts in blanks
and Cu-samples. For \Uranium and \tracerU measurements,
the isobaric interference can be ignored since there is
no isobar for these two isotopes. The polyatomic interference
is analyzed below.

A polyatomic ion can be formed in the plasma, and it can mimic
a monatomic ion with the same mass-to-charge ratio.
In the following, $C_{m/z}$ are denoted as the measured
counts per second with ICP-MS at a particular
mass-to-charge value $m/z$.
The ubiquitous types of polyatomic ions,
such as MO$^+$, MOH$^+$, MO$^{2+}$, MO$_2$H$^+$, M$^{2+}$,
M$_2$O$^+$, and MAr$^+$, were taken into account,
and such a list could be not complete depending on the
possible presence of other elements.
The possible polyatomic ions that contribute $C_{233}$ are
$^{232}$ThH$^+$, $^{201}$Hg$^{16}$O$^{2+}$,
$^{200}$Hg$^{16}$O$_2$H$^+$, and $^{193}$Ir$^{40}$Ar$^+$.
The concentration of \Thorium, $^{200}$Hg, $^{201}$Hg
and $^{193}$Ir in Cu-samples was measured to be approximately
10$^{-12}$ g/g level or even lower; thus, the
probability of forming these polyatomic ions is quite low.
The Cu-samples and blanks without adding \tracerU were tested
to check the interference, and the measured $C_{233}$ was
3.0$\pm$0.0, 2.1$\pm$0.9, and 0.2$\pm$0.2 for Cu-samples,
blanks, and 5\% \nitric, respectively.

The polyatomic interference on $C_{238}$ cannot be measured
directly because \Uranium is naturally existing and widely
distributed. Instead, we divided the possible interference into
three groups and theoretically calculated their contributions.
Dedicated tests were performed to verify the calculations;
the details appear below.

\begin{itemize}
\item Group A: $^{119}$Sn$_2^+$, $^{198}$Hg$^{40}$Ar$^+$, $^{111}$Cd$_2$$^{16}$O$^+$, $^{206}$Pb$^{16}$O$_2^+$

If any of these polyatomic ions is generated, there will be also
contributions to $C_{240}$ if the heaviest atom is replaced
with its natural isotope, as shown in Table~\ref{tab:abundance}.
Because the element with $m/z=240$ does not naturally exist,
non-zero $C_{240}$ observed by ICP-MS should come from the
polyatomic interference. For each polyatomic ion in group A,
the ratio of $C_{240}$ to $C_{238}$ is calculated according
to the natural abundances.
Table~\ref{tab:measured} shows the measured $C_{m/z}$ with ICP-MS.
Taking the maximum $C_{240}$ and the minimum ratio in
Table~\ref{tab:abundance}, the interference on $C_{238}$ can be
conservatively estimated to be $<0.01$ ppt.

\item Group B: $^{198}$Pt$^{40}$Ar$^+$

Similar estimation can be applied for $^{198}$Pt$^{40}$Ar$^+$.
If PtAr$^+$ is generated, the ratios $C_{234}$($^{194}$Pt$^{40}$Ar$^+$)/$C_{238}$($^{198}$Pt$^{40}$Ar$^+$) and $C_{235}$($^{195}$Pt$^{40}$Ar$^+$)/$C_{238}$($^{198}$Pt$^{40}$Ar$^+$)
are calculated to be 4.6:1 and 4.7:1, respectively.
Using the measured $C_{234}$ and $C_{235}$ in Table~\ref{tab:measured},
the projected contribution of $^{198}$Pt$^{40}$Ar$^+$ to $C_{238}$
would also be $<0.01$ ppt.

\item Group C: $^{205}$Tl$^{16}$O$_2$H$^+$

$^{205}$Tl$^{16}$O$_2$H$^+$ is a polyatomic ion with four atoms.
The $^{205}$Tl concentration in a blank or Cu-sample is low,
and thus the formation of TlO$^+$ is rare, as shown in the lower part
of Table~\ref{tab:measured}, where the measured $C_{205}$,
$C_{219}$, and $C_{221}$ are corresponding to $^{205}$Tl,
$^{203}$Tl$^{16}$O$^+$, and $^{205}$Tl$^{16}$O$^+$, respectively.
The possibility of forming four-atomic ions such as
$^{205}$Tl$^{16}$O$_2$H$^+$ is much lower than that of forming TlO$^+$,
and thus it can be neglected in this assay.
\end{itemize}

In conclusion, the polyatomic interference on \tracerU and
\Uranium measurements for both blanks and Cu-samples are
less than 0.01 ppt. If the uranium in copper
is at the 0.1--1-ppt level, such sources of interference
can be ignored. Furthermore, they can be effectively
subtracted using the blank measurements.

\begin{table}[!htb]
\centering
\caption{Abundance of metal ions and calculated $C_{240}$/$C_{238}$ ratio. \label{tab:abundance}}
\begin{tabular}{cccc}
\hline
Polyatomic ions & Abundance & $C_{240}$/$C_{238}$\\
\hline
Sn$_2^+$ & \makecell*[l]{$^{119}$Sn = 8.58\%\\$^{120}$Sn = 32.85\%} & 14.7:1 \\
\hline
Hg$^{40}$Ar$^+$ & \makecell*[l]{$^{198}$Hg = 10.02\%\\$^{200}$Hg = 23.13\%} & 2.3:1 \\
\hline
Cd$_2^{16}$O$^+$ & \makecell*[l]{$^{111}$Cd = 12.75\%\\$^{112}$Cd = 24.0\%} & 1.9:1 \\
\hline
Pb$^{16}$O$_2^+$ & \makecell*[l]{$^{206}$Pb = 24.1\%\\$^{208}$Pb = 52.3\%} & 2.2:1 \\
\hline
\end{tabular}
\end{table}

\begin{table}[!htb]
\centering
\caption{Measured counts per second at different $m/z$ with ICP-MS.}
\label{tab:measured}
\begin{tabular}{cccc}
\hline
Measured $C_{m/z}$  & 5\% \nitric & Blanks & Cu-samples\\
\hline
$C_{240}$ & 0.4$\pm$0.3 & 4.4$\pm$2.5 & 3.0$\pm$1.8\\
\hline
\hline
$C_{234}$ & 0.2$\pm$0.2 & 9.6$\pm$4.4 & 6.3$\pm$3.2\\
$C_{235}$ & 1.6$\pm$2.3 & 6.8$\pm$1.9 & 5.6$\pm$2.5\\
\hline
\hline
$C_{205}$($^{205}$Tl) & - & 88$\pm$60 & 144$\pm$33\\
$C_{219}$($^{203}$Tl$^{16}$O$^+$) & - & 0$\pm$0 & 5$\pm$5\\
$C_{221}$($^{205}$Tl$^{16}$O$^+$) & - & 5$\pm$0 & 8$\pm$6\\
\hline
\end{tabular}
\end{table}

\subsection{\Uranium contaminations from reagents}

Table~\ref{tab:purity} summarizes the analysis of \Uranium
contamination from various sources,
and lists the required amount of reagents to deal with 3.2~g copper.
According to the used mass and the measured \Uranium content
of each reagent, ammonia water contributed the largest contamination,
and the \tracerU standard contributed the least.
\ZrCl had the worst purity, but its contribution
was not significant due to the tiny amount used.
Environmental contamination, due to dust in the air and
surface impurities from labware, was estimated by subtracting
the reagents' contribution from the total.
Although not listed in Table~\ref{tab:purity}, the \Thorium
contamination of nitric acid, ammonia water, and \ZrCl were measured
to be $<$0.01, $\sim$0.02, and $\sim$150 ppt, respectively.
For future prospects, we expect to further purify the ammonia
water and \ZrCl and reduce the impurities by a factor of
10 and 100, respectively. Further distillation to nitric acid may
result in a factor-of-2 improvement.
The environmental contamination will be reduced by
at least an order of magnitude with a new class 100 clean laboratory
being built. The total contamination after these improvements is
projected to be within 0.06 ppt, which will significantly improve
the detection limits.

\begin{table}[!htb]
\centering
\caption{Analysis of \Uranium contamination. For each reagent,
measured \Uranium contamination and required amount to deal
with 3.2~g copper is listed. Fractions represent relative
contributions of each source in this assay. Significant improvement
can be expected with future purifications.}
\label{tab:purity}
\begin{tabular}{cccc}
\hline
& \multicolumn{2}{c}{\Uranium contamination (ppt)}\\
\cline{2-3}
\raisebox{2.3ex}[0pt]{Sources} & this assay &  projection\\
\hline
\nitric (67\%, 18.4 g) & $<$0.01 $^{\rm a}$ (16\%) & 0.005 \\
\ammoniaW (25\%, 19.5 g) & $\sim$0.034 (59\%) & 0.003 \\
\ZrCl (0.001 g) & $\sim$130 (12\%) & 1 \\
\tracerU standard (1 g) & $\sim$0.014 (1\%) & 0.014 \\
\\
environmental and labware & 0.04 (12\%) & 0.004\\
\\
total & 0.35 & 0.056\\
\hline
\end{tabular}
\begin{tabnote}
  $^{\rm a}$ Quoted from OPTIMA specification.
\end{tabnote}
\end{table}

\section{Results and discussion}
\label{sec:result}

The method detection limit (MDL) is defined as the minimum measured
concentration of a substance that can be reported with 99\% confidence
so that the measured concentration is distinguishable from the method
blank results.
In this work, the MDL was obtained by following the U.S. Environmental Protection Agency (EPA) process~\cite{AppendixB}. The blanks with
a matrix similar to that of the Cu-samples were prepared by following
the procedures in Sec.~\ref{sec:samplePrep}, and then measured with
two independent methods as a cross-check, namely
the isotope dilution method (IDM) and standard addition method (SAM).
As discussed in Sec.~\ref{sec:recovery}, \tracerU was used to calibrate the
recovery efficiency during pre-treatment, and the SAM and IDM shared the
same premise that the recovery of \tracerU and \Uranium are the same.

\subsection{Isotope dilution method}

Isotope dilution analysis is known as an important analytical technique
for the quantification of mass-spectrometric data~\cite{Becker2012}.
It is a relative approach and only involves the measurement of isotopes
of the same element, thus eliminating differences in chemical behavior.
If the isotopes are mixed homogeneously, their ratio is expected not to
change during the entire analytical procedure,
including sample preparation, analyte separation, and enrichment.
In addition, the matrix effect and instability of ICP-MS have exactly
the same influence on each isotope; thus, they have negligible impact on
the result. In this work, a known quantity of Cu-\nitric solution
(for Cu-sample) or corresponding \nitric (for blank) was mixed
homogeneously with a known amount of \tracerU standard,
and several parallel samples were prepared and measured.

For isotope dilution analysis, the measured signal ratio of \Uranium
to \tracerU was used to calculate the \Uranium introduced during
the pre-treatment of each gram of Cu:
\begin{equation*}
R = (n_{238}+n'_{238} ) / n'_{233},
\end{equation*}
where $R$ is the measured ratio of \Uranium to \tracerU, $n_{238}$
is the total \Uranium not originating from the tracer solution, and
$n'_{238}$ and $n'_{233}$ are the total \Uranium and \tracerU from
the tracer solution, respectively. Note that \tracerU does not exist
naturally, so it does not appear in the denominator.
Given the measured $R$ and the known $n'_{233}$ and $n'_{238}$,
$n_{238}$ can be easily calculated.
In this assay, for Cu-samples, $n_{238}$ was the total amount of \Uranium
originating from Cu and introduced during the pre-treatment,
while, for blanks, $n_{238}$ was only the amount of \Uranium introduced
during the pre-treatment.

The isotope dilution method measures the ratio of \Uranium to \tracerU
in blanks or Cu-samples, and thus it is necessary to quantify the
recovery efficiency of \tracerU because it is an important indicator
used to evaluate the effectiveness of the pre-treatment method.
The standard addition method was chosen to quantify the recovery
efficiency, instead of the commonly used external standard method
due to the possible matrix effects mentioned in
Sec.~\ref{sec:specInterference}.

\subsection{Standard addition method}

As discussed in Sec.~\ref{sec:specInterference}, the Cu-samples or blanks
in this assay contained certain impurities that may cause a matrix effect.
If using the external standard method, the calibration curve built with
pure standard solutions may result in incorrect data. Thus,
the standard addition method (also see~\cite{Becker2012}),
as one of the internal standard methods,
was applied to solve the matrix effect by adding the standard to
the samples to match the matrix.

\begin{figure}[!htb]
\centering
\includegraphics[width=0.6\columnwidth]{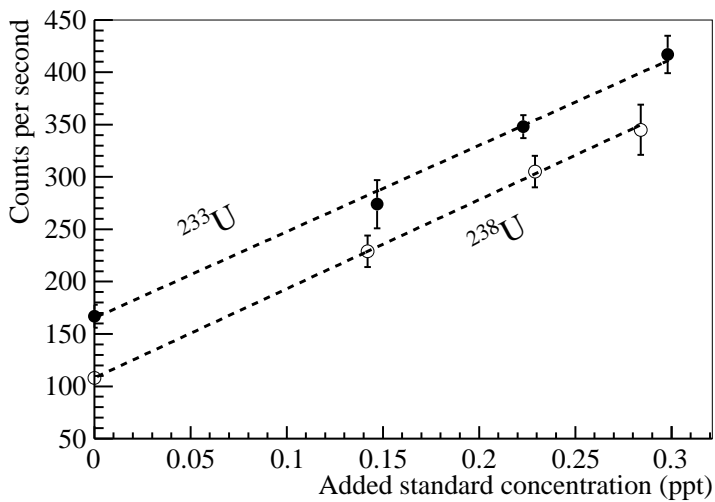}
\caption{Measured signals vs different \tracerU (solid circles) and \Uranium (open circles) concentrations in one of the blanks by using the standard addition method; error bars are standard deviations of parallel samples. Linear function fits showed good linearities for both \tracerU and \Uranium. } \label{fig:measureSAD}
\end{figure}

A series of \tracerU and \Uranium standards with different
concentrations was prepared in advance. The blank or Cu-sample after
pre-treatment was equally divided into four or five sub-samples,
typically 1 g for each sub-sample.
Then, 0.2-g standard solutions with different concentrations of
\tracerU and \Uranium were added to the parallel sub-samples of
blanks or Cu-samples.
The final mixtures, namely the test samples, were injected into
the ICP-MS instrument for analysis.
Figure~\ref{fig:measureSAD} shows an example of the blank data.
A linear function was used to fit the measured
signals (in counts-per-second unit) versus different added
\tracerU or \Uranium contents in the test samples, respectively,
and the ratio of the intercept to the slope represented the
concentration of \tracerU or \Uranium in the first test sample
in which the concentration of added \tracerU or \Uranium was zero.
Therefore, the concentrations of \tracerU and \Uranium in the blanks
could be obtained.
The recovery efficiency of \tracerU was determined and used to
extrapolate the overall \Uranium contamination when dealing
each gram of raw Cu.

\subsection{MDL results}

\begin{figure}[!htb]
\centering
\includegraphics[width=0.6\textwidth]{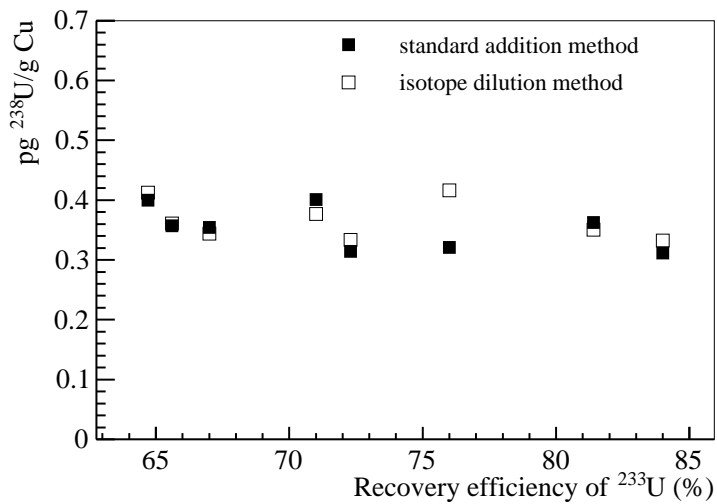}
\caption{Measured \Uranium vs the different \tracerU recovery
efficiencies. No obvious dependency was found. } \label{fig:MDL}
\end{figure}

Eight re-duplicative blanks were processed and measured, and
Table~\ref{tab:MDL} shows that a method detection limit of $\sim$0.1 ppt
is consistently achievable for the two different quantitative methods.
The measured \Uranium contents versus the \tracerU recovery
efficiencies are shown in Fig.~\ref{fig:MDL},
and no obvious dependency was observed, indicating the robustness of
the pre-treatment and analysis.

\begin{table}[!htb]
\centering
\caption{Method detection limit (MDL) for \Uranium
in copper obtained by measuring eight re-duplicative blanks.
SD denotes standard deviation, and MDL is calculated
as 2.998$\cdot$SD~\cite{AppendixB}. Two quantitative methods,
standard addition method (SAM) and isotope dilution method (IDM),
were used to cross-check each other. \tracerU recovery
efficiencies obtained by SAM are also listed.}
\label{tab:MDL}
\begin{tabular}{cccc}
\hline
\makecell{Blank\\No.} & \makecell{meas. by SAM\\(ppt)} & \makecell{meas. by IDM\\(ppt)} & \makecell{$\epsilon(\rm{^{233} U})$}\\
\hline
1 &	0.3206 & 0.4162 & 76.0\%\\
2 &	0.4010 & 0.3768 & 71.0\%\\
3 &	0.3543 & 0.3438 & 67.0\%\\
4 &	0.3116 & 0.3325 & 84.0\%\\
5 &	0.3623 & 0.3505 & 81.0\%\\
6 &	0.3143 & 0.3333 & 72.3\%\\
7 &	0.3572 & 0.3608 & 65.6\%\\
8 &	0.4001 & 0.4125 & 64.7\%\\
\hline
average	& 0.3527 & 0.3658 & 72.2\%\\
SD & 0.0356 & 0.0333 & - \\
MDL & 0.107 & 0.099 & - \\
\hline
\end{tabular}
\end{table}

\section{Summary}
\label{sec:summary}

A novel co-precipitation method was developed to determine the
trace or ultra-trace amount of \Uranium in high-purity copper,
and the current MDL could achieve 0.1 ppt.
The matrix effect could be effectively suppressed by using the isotope
dilution method or standard addition method. The interference for
ICP-MS was evaluated to be lower than 0.01 ppt and thus could be
ignored for this assay.
The \Thorium in copper could be determined by using exactly the
same method and $^{229}$Th as the tracer. Preliminary tests
achieved a similar MDL and recovery efficiency for \Thorium.

Several efforts are in progress to further improve the MDL.
The cleanness of the pre-treatment lab will be improved to class 100.
A more rigorous validation procedure is being developed to verify the
surface cleanness of the vessels and filter units.
Further purification of ammonia water has been undertaken and preliminary
tests have shown that the purity can be improved by an order of magnitude.
Further purification of \ZrCl should be able to realize at least
a 2-order-of-magnitude improvement by using an ion exchange resin
column or co-precipitation.
An alternative option is to find other co-precipitators that can have
higher purity or can be purified easily.
Nevertheless, along with the above improvements, the sources of
polyatomic interference will become a major factor that affects MDL.
The collision cell incorporated into the ICP-MS setup can be
used to suppress such effects.
Collision reaction cell technology is one of the major breakthroughs
to obviate the polyatomic or isobaric interference for ICP-MS measurement
based on ion-molecule chemistry. The ions entering the multipole
(quadrupole, hexapole or octupole) system will collide or react with
the collision gas, such as H$_2$ or He. Then, the polyatomic ions can be
changed into interference-free substances, or the elements to be
measured can be turned into other ions that will not cause interference.
Ultimately, the proposed method promises to improve the MDL of \Uranium
and \Thorium in copper by an order of magnitude. Methods determining
the \Uranium and \Thorium in, e.g., other metals, quartz, and silicon,
will be developed in the future as well.

\section*{Acknowledgment}
This work was supported by National Natural Science Foundation of China
(Grant No.11820101005), Strategic Priority Research Program of the Chinese
Academy of Sciences (Grant No. XDA10010500) and
Youth Innovation Promotion Association of CAS (2014008).


\end{document}